\documentclass[prl,aps,amssymb,twocolumn,10pt,floatfix,showpacs,superscriptaddress]{revtex4-1}

\usepackage{color,graphicx}
\usepackage{bm}

\usepackage{amsmath}
\usepackage{amsfonts}
\usepackage{amssymb}

\usepackage{subfigure}
\usepackage[latin1]{inputenc}

\usepackage{pifont}

\newcommand{\ket}[1]{\ensuremath{\left| #1 \right\rangle}}

\newcommand{\ketbra}[2]{\ensuremath{\left|#1 \right\rangle\left\langle #2 \right|}}

\renewcommand{\-}{\,-\,}

\newcommand{\br}{\mathbf{r}}

\newcommand{\Tr}{\ensuremath{\mathrm{Tr}}}

\let\oldmarginpar\marginpar
\renewcommand\marginpar[1]{\-\oldmarginpar[\raggedleft\tiny #1]%
{\raggedright\tiny #1}}

\begin{document}
\title{Creating a bosonic fractional quantum Hall state by pairing fermions}

\author{C\'ecile Repellin}
\email{repellin@pks.mpg.de}
\address{Max-Planck-Institut f\"ur Physik komplexer Systeme, N\"othnitzer Str. 38, 01187 Dresden, Germany} 

\author{Tarik Yefsah}
\email{tarik.yefsah@lkb.ens.fr}
\address{Laboratoire Kastler Brossel, ENS-PSL Research University, CNRS, UPMC-Sorbonne Universit\'es and Coll\`ege de France, 75005 Paris, France}

\author{Antoine Sterdyniak}
\email{antoine.sterdyniak@mpq.mpg.de}
\address{Max-Planck-Institute of Quantum Optics, Hans-Kopfermann-Str. 1, D-85748 Garching, Germany}


\begin{abstract}
We numerically study the behavior of spin--$1/2$ fermions on a two-dimensional square lattice subject to a uniform magnetic field, where opposite spins interact via an on-site attractive interaction. Starting from the non-interacting case where each spin population is prepared in a quantum Hall state with unity filling, we follow the evolution of the system as the interaction strength is increased. Above a critical value and for sufficiently low flux density, we observe the emergence of a twofold quasidegeneracy accompanied by the opening of an energy gap to the third level. Analysis of the entanglement spectra shows that the gapped ground state is the bosonic $1/2$ Laughlin state. Our work therefore provides compelling evidence of a topological phase transition from the fermionic quantum Hall state at unity filling to the bosonic Laughlin state at a critical attraction strength.
\end{abstract}
\date{\today}
\maketitle

Topological phases of matter represent a subject of intense research~\cite{Hasan2010review, Qi2011, Nobel2016}, offering the prospect of realizing fault tolerant quantum computation~\cite{Kitaev2003, Hasan2010review, Qi2011}. The rapid growth of this field, since the first observation of the quantum Hall effect, is largely due to experimental progress in producing high quality and purity materials~\cite{Hasan2010review, Qi2011}. In this context, ultracold quantum gases, which feature clean, flexible and well characterized environments, appear as promising systems. Furthermore, as they are comprised of electrically neutral particles that can be either fermionic or bosonic, atomic quantum gases may provide a new vista on the subject~\cite{Dalibard2011,Dalibard2015,Goldman2016,Cooper2008review}. Notable achievements in this direction include the creation of Bose-Einstein condensates (BEC) in the lowest Landau level via fast rotation~\cite{Schweikhard2004,Bretin2004}, the realization of the Harper--Hofstadter~\cite{miyake2013realizing,Aidelsburger2013,Aidelsburger2015} and Haldane~\cite{jotzu2014experimental} models in optical lattices, and more recently the observation of chiral currents in atomic ladders~\cite{Atala2014, Stuhl2015, Mancini1510}, as well as the measurement of the second Chern number of a non-Abelian Yang monopole~\cite{2016arXiv161006228S}. 

A remarkable feature of ultracold gases is the ability to use tunable attractive interactions to convert, in real time, a pair of distinguishable fermions into a tightly bound bosonic molecule (BM)~\cite{Zwerger2011BECBCS, Zwierlein2013book}. From the perpective of topological matter, this feature opens unique possibilities. Suppose a two-dimensional (2D) spin-$1/2$ fermionic system is initially prepared in an integer quantum Hall (IQH) state for each spin component ($\uparrow$ and $\downarrow$), with filling factor $\nu_\uparrow=\nu_\downarrow=n$. As the attraction strength is brought to the strong binding limit, each fermion pair forms a bosonic molecule that carries twice the fermion (neutral) charge, and hence experiences twice the magnetic flux seen by a fermion. As a result, the final many-body system would be characterized by a filling factor $\nu_{\uparrow,\downarrow}/2$, provided all bosons occupy the lowest band. 

Building upon this observation, K.~Yang and H.~Zhai pointed out, almost a decade ago, the possibility of using a Feshbach resonance to drive a transition between a fermionic IQH state and a bosonic fractional quantum Hall (FQH) Laughlin state~\cite{PhysRevLett.100.030404}. These two topologically distinct phases were shown to be separated by a quantum phase transition whose critical behavior remains an important open question~\cite{Wen1993,Chen1993,Pryadko1994}. More recently, T-L.~Ho proposed the use of a rapid sweep through this transition to \textit{project} fermionic IQH states onto the BEC side of the Feshbach resonance in order to reveal the bosonic FQH structure of its center-of-mass wave function~\cite{2016arXiv160800074H}. However, so far, these two analytic studies have not been supported by a microscopic approach. It is also important to determine whether such transition can occur in the presence of a lattice, which can strongly alter the system's topological properties~\cite{Peierls1933,Harper1955,PhysRevLett.106.236804,2011NatCo...2E.389S,PhysRevB.14.2239,PhysRevLett.106.236802,PhysRevX.1.021014,parameswaran2013fractional,bergholtz2013topological,wang2012fractional,PhysRevLett.109.186805,sterdyniak2013series, moller-PhysRevLett.115.126401}.

In this Letter we numerically study the behavior of spin-$1/2$ fermions on a 2D lattice subject to a homogeneous magnetic field, with attractive on-site interaction between $\uparrow$ and $\downarrow$ spins. Using energy and particle entanglement spectroscopy on finite size systems, we provide evidence for the existence of a topological phase transition from the fermionic IQH $\nu_{\uparrow,\downarrow}=1$ state to the $1/2$ bosonic Laughlin state above a critical attraction strength. As the transition occurs in the low flux limit, our findings also serve as a microscopic support to the quantum field description~\cite{PhysRevLett.100.030404} and the wave function analysis~\cite{2016arXiv160800074H} suggested in the continuum case.

\begin{figure}[t!]
\includegraphics[width=86mm]{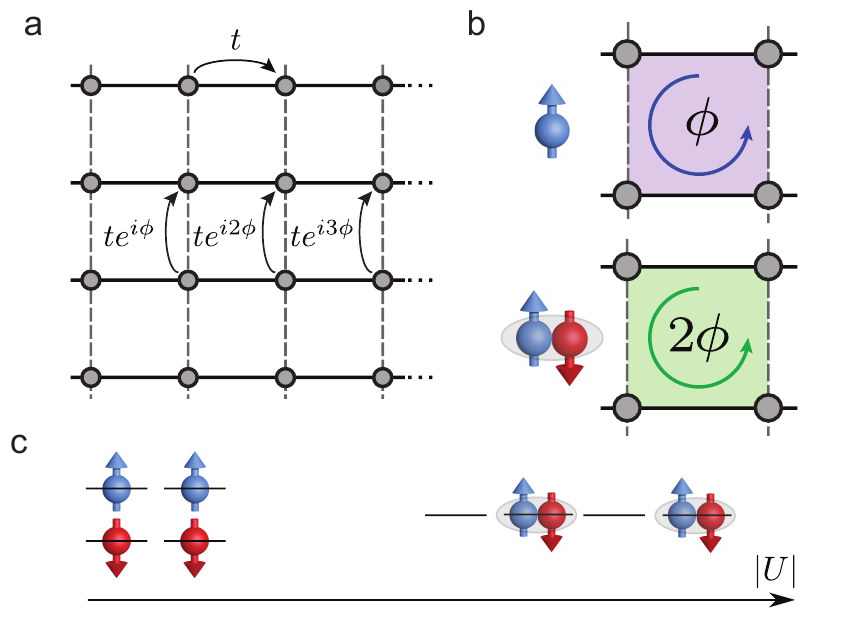}
\caption{(a) Schematic description of the Harper--Hofstadter model on a square lattice with Landau gauge in the $y$-direction. Each plaquette is pierced by a flux $\phi = 2\pi \alpha$, where $\alpha$ is the flux density defined in the main text. The phase $\phi$ is proportional to $\varphi_{\br\br'}$ in Eq.~\ref{eq:Hamiltonian}. For simplicity, we omit the minus sign in the tunneling. (b) The flux seen by a bosonic molecule is twice the one of a single fermion. (c) Possible scenario of a phase transition induced by an attractive on-site interaction between the fermions. At small $|U|$, the system is described by the $\nu_{\uparrow,\downarrow}=1$ IQH state. At large $|U|$, fermions bind into bosonic molecules and the system forms the $1/2$ Laughlin state.} 
\label{fig::Schema}
\end{figure}

The system of interest here is described by the Fermi-Hubbard model with minimal coupling to a gauge field via Peierl's substitution. The Hamiltonian reads
\begin{equation}
\label{eq:Hamiltonian}
\mathcal{H} =  - t \sum_{\langle \br,\br' \rangle,\sigma=\uparrow,\downarrow} \left( e^{i \varphi_{\br\br'}} c_{\br,\sigma}^{\dagger} c_{\br',\sigma} + h.c. \right)  
  + U \sum_{\br}  n_{\br,\uparrow} n_{\br,\downarrow}
  \end{equation}
where $\langle \br,\br' \rangle$ denotes neighboring lattice sites $\br=(x,y)$, $c_{\br,\sigma}$ (resp. $c_{\br,\sigma}^{\dagger}$)  annihilates (resp. creates) a fermion with spin $\sigma$ at site $\br$, $n_{\br,\sigma} = c_{\br,\sigma}^{\dagger}c_{\br,\sigma}$ and $\varphi_{\br\br'} =\int_\br^{\br'}\mathbf{A}\cdot\mathrm{d}\mathbf{l}$ are Aharonov-Bohm phases derived from the coupling to the underlying vector potential $\mathbf A$ (see Fig.~\ref{fig::Schema}a). Fermions of opposite spin interact through an attractive on-site interaction of strength $U \leq 0$. In the absence of interactions ($U = 0$), the spectrum of this model is the fractal Hofstadter butterfly~\cite{PhysRevB.14.2239}. 

We consider $2N$ fermions ($N$ per spin state) distributed over $N_{\rm s}$ sites of the square lattice and experiencing a total magnetic flux $N_{\phi}\,\Phi_0$, where $\Phi_0$ is the flux quantum. We define the flux density $\alpha = \frac{N_{\phi}}{N_{\rm s}}$, meaning that the flux per lattice site is $\alpha\Phi_0$. We choose $\alpha=1/q$ with $q$ an integer, such that each magnetic unit cell is composed of $q$ lattice sites, and is characterized by a single-particle spectrum with $q$ bands. The filling factor per spin state of the lowest band therefore reads $\nu_{\uparrow,\downarrow} = N/N_{\phi}$.  We allow the interaction strength $|U|$ to take arbitrarily large values and therefore take into account all $q$  bands. While this restricts the numerically accessible system sizes to $2N\leq8$, it has the crucial advantage of including band mixing and band dispersion effects, thus capturing all features of the Harper--Hofstadter model. Our aim is to study the nature of the many-body ground state of the system as a function of flux density and interaction strength. We focus on the situation where the number of fermions per spin state is equal to the number of flux quanta ($N=N_\phi$), such that the filling factor per spin state is $\nu_{\uparrow,\downarrow}=1$. 

In order to build an intuition about the system's possible behavior, let us consider first the low flux limit $\alpha \ll1$, where the Harper--Hofstadter model is similar to the  continuum case. There, one can anticipate that a system initially prepared in the $\nu_{\uparrow,\downarrow}=1$ IQH state could evolve into the bosonic $1/2$ Laughlin state~\cite{PhysRevLett.100.030404}. Indeed, for increasing attraction, $\uparrow$ and $\downarrow$ fermions will form pairs of increasing binding energy and decreasing pair size. For sufficiently large attraction, each fermion pair, whose size becomes smaller than any other length scale, acts as a bosonic molecule (BM) experiencing a flux density $\tilde{\alpha} = 2\alpha$. Therefore the emerging many-body system -- composed of $N$ bosonic molecules -- could feature a filling factor $\tilde{\nu} =\frac{1}{\tilde{\alpha}}\frac{N}{N_{\rm s} }= \frac{\nu_{\uparrow,\downarrow}}{2}=1/2$ of the lowest band. This scenario, depicted in Fig.~\ref{fig::Schema}, remains to be revealed via a microscopic description in both the continuum and lattice systems.

In lattice systems, it was shown that the $1/2$ bosonic Laughlin state may form in the Harper--Hofstadter model at low flux ($\tilde{\alpha} \ll 1$) for both hard-core and finite repulsive interactions~\cite{PhysRevLett.94.086803, PhysRevA.76.023613}. Therefore, in the strong binding $|U|/t\gg 1$ and low flux $\tilde{\alpha} \ll 1$ limits, the Laughlin state represents a reasonable candidate ground state of the Hamiltonian (Eq.~\ref{eq:Hamiltonian}). However, our model comes with an additional difficulty: the composite nature of the bosons is associated with the energy scale $|U|$, whose competition with the other energy scales could favor competing phases. In the following, we describe our analysis of the microscopic Hamiltonian (Eq.~\ref{eq:Hamiltonian}).

\paragraph{Energy spectra ---}

\begin{figure}[t!]
\includegraphics[width=86mm]{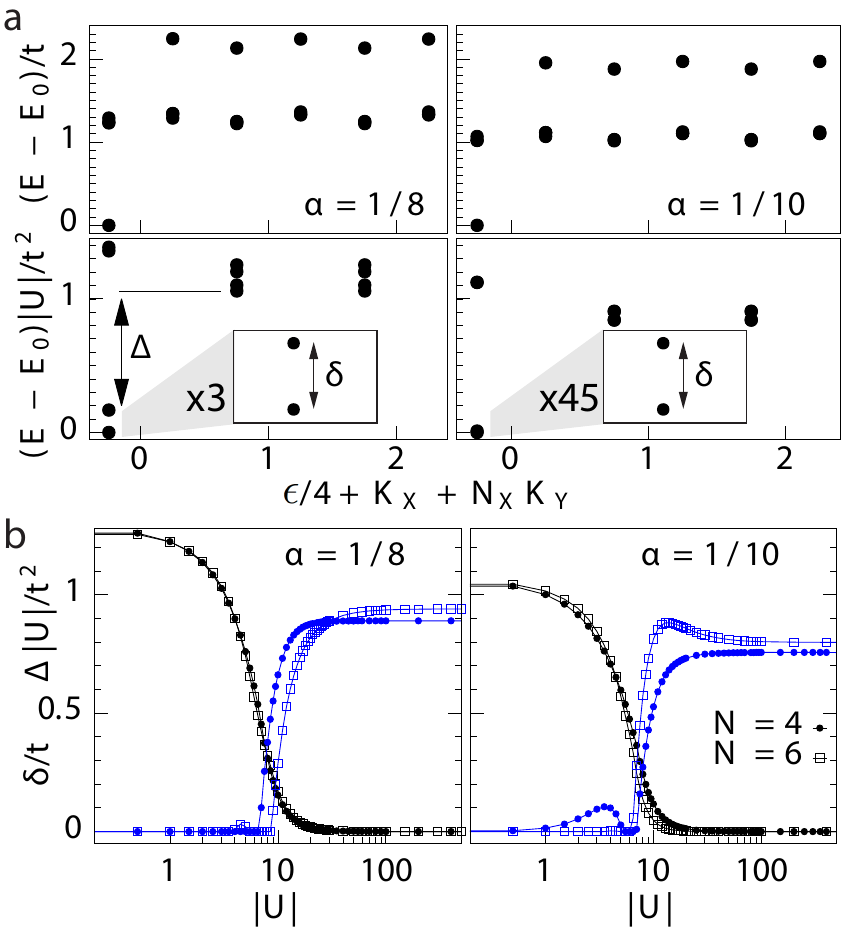}
\caption{(a) Energy spectrum of $2N=6$ fermions for flux densities $\alpha=1/8$ (left column) and $\alpha=1/10$ (right column) and different interaction strengths $U=-t$ (upper panel) and $U=-30t$ (lower panel). The ground state is not degenerate for $U=-t$ while it is quasi twofold degenerate for $U=-30t$. The even states under spin inversion have an energy offset $|U|$ and, thus, are not visible on the graphs. (b) Evolution of the first ($\delta$) and second ($\Delta$) energy gaps with increasing interaction strength $|U|$.}
\label{fig::Energy}
\end{figure}

For our numerical calculations, we apply periodic boundary conditions in both directions and choose the number of sites in each direction such that the aspect ratio is close to one. The momenta $(K_x, K_y)$ are defined by the translations of the magnetic unit cell on the lattice~\cite{PhysRevB.48.8890}. The Hamiltonian matrix is block-diagonal in these quantum numbers as well as in the total magnetization $S_z$ and spin inversion parity $\epsilon$ for $S_z = 0$.
We have performed the exact diagonalization of the Hamiltonian of Eq.~\eqref{eq:Hamiltonian} at $\nu_{\uparrow,\downarrow} = 1$ up to $q = 10$ using these symmetries.
Due to computational constraints, the number of fermions is limited to $2N = 8$. Our results are plotted in Fig.~\ref{fig::Energy} for $\alpha=1/8$ and $\alpha=1/10$.
When $U = 0$, the fermions completely fill up the lowest band of the Harper--Hofstadter model and realize the  $\nu_{\uparrow,\downarrow} = 1$ IQH state. It is characterized by a single ground state in the $(K_x, K_y) = (0,0)$ sector separated by a large gap $\delta$ from higher energy states. As shown in Fig.~\ref{fig::Energy}a (top panel), a moderate interaction ($U = -t$)  does not change this picture. On the contrary, when $|U|/t = 30$ (Fig.~\ref{fig::Energy}a, bottom panel), the gap $\delta$ is small compared to the energy difference between the second and third lowest energy eigenstates $\Delta$. This quasi twofold degeneracy is characteristic of the $\tilde{\nu} = 1/2$ Laughlin state on a torus (here the lattice with periodic boundary conditions). The quasidegenerate ground states are found in the $(K_x, K_y) = (0,0)$, $S_z = 0$ sector and have a parity $\epsilon = (-1)^{N}$ under spin inversion, which is consistent with this interpretation. We computed these spectra for various interaction strengths $U$ and the results are summarized in Fig.~\ref{fig::Energy}b. The gap $\delta$ is seen to persist for attractive interactions as strong as $U \simeq -10t$, regime at which $\Delta$ starts to increase. $\Delta|U|$ eventually saturates at $|U|/t \simeq 30$. Note that for $\alpha>1/8$ we do not observe any twofold quasidegeneracy at any value of $U$, probably due to larger lattice effects. This is reminiscent of the bosonic Harper--Hofstadter calculations, where the Laughlin state is not observed for $\tilde{\alpha}\gtrsim 1/3$ \cite{PhysRevLett.94.086803,PhysRevA.76.023613}. Interestingly, the closing of the IQH gap $\delta$ and opening of the second gap $\Delta$ seem to coincide for all system sizes considered here, and the critical value of $U$ shows very little dependence with system size and flux density. This suggests a direct phase transition between the IQH and the Laughlin phase.

As complementary evidence to the ground state degeneracy, we also checked the degeneracy of the quasihole excitations~\cite{suppmatCP}. However, note that in a finite system a charge density wave (CDW) can have the same low energy spectrum as the Laughlin state (and likewise concerning their respective excitations). The precise nature of the phase can be determined using entanglement spectroscopy~\cite{PhysRevLett.101.010504,PhysRevLett.106.100405}, which is the purpose of the following section.

\paragraph{Entanglement spectroscopy ---}

In order to establish the nature of the twofold ground state at large $|U|$, we use the particle entanglement spectrum (PES)~\cite{PhysRevLett.106.100405}. The entanglement spectrum was originally introduced to extract the edge spectrum from the ground state wave function~\cite{PhysRevLett.101.010504}. The PES uses a similar method to reveal the nature of bulk quasihole excitations and was crucial to fully establish the emergence of FQH states in Chern insulators~\cite{PhysRevX.1.021014}. Unlike the number of quasihole states in the energy spectrum, it can distinguish between FQH and CDW states in Chern insulators~\cite{2012arXiv1204.5682B} and identify superfluid phases in the Hofstadter model~\cite{PhysRevB.86.165314}. The PES is the spectrum of $-\log \rho_A$ where $\rho_A = \Tr_B \rho$ is the reduced density matrix obtained by tracing over $N_B \equiv 2N - N_A$ particles, the labels $A$ and $B$ referring to two complementary parts of the whole system. Thus, $\rho_A$ commutes with the total momentum $(K_{x,A},K_{y,A})$ and the spin $\mathbf{S}_{A}$ of the subsystem $A$, as well as the spin inversion parity $\epsilon_A$ when $S_{z,A} =0$. When the ground state is almost twofold degenerate, we consider $\rho = \frac{1}{2}\sum_{i=1}^2 \ketbra{\Psi_i}{\Psi_i}$ where $\ket{\Psi_1}$, $\ket{\Psi_2}$ are the two ground states. Generically, the PES has low entanglement energy levels separated from higher entanglement levels by an entanglement gap, which is infinite for model FQH states, but finite in the Hofstadter model~\cite{PhysRevB.86.165314}. The number of levels below the entanglement gap is related to the number of quasihole states and is a fingerprint of a given topological phase.

\begin{figure}[t!]
\includegraphics[width = 86mm]{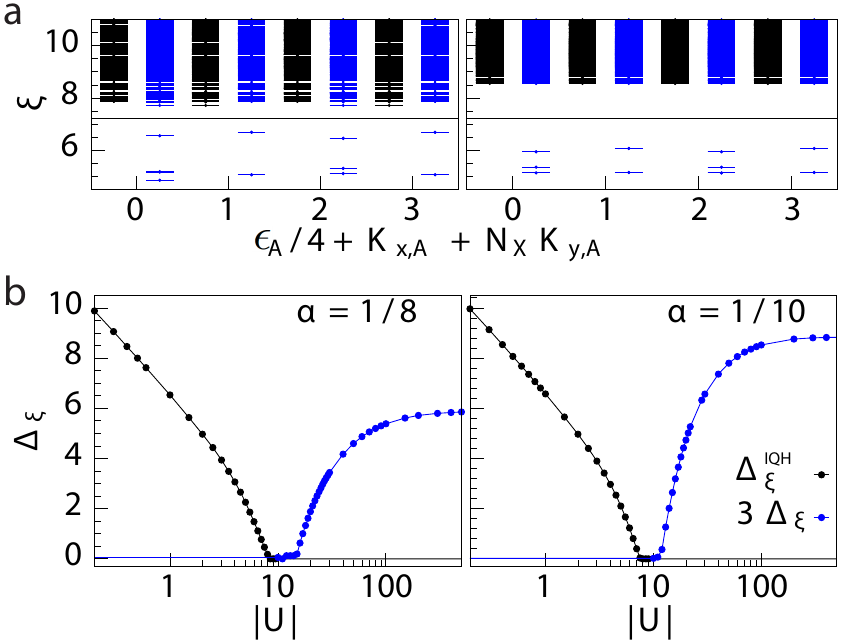}
\caption{(a) Entanglement spectrum for $2N=8$ fermions, $|U|=30t$ and $\alpha = 1/8$ (left) and $\alpha = 1/10$ (right), with a particle partition $N_A = 4$ in the sector $S_z^A = 0$. The low entanglement energy levels are all even (blue) under spin inversion which reveals the existence of a pairing gap. The horizontal black line indicates the midgap energy. The number of states below this line is 20 (6,4,6,4), which is the expected counting for the bosonic Laughlin state. (b) Evolution of the entanglement gap of the IQH ($\Delta^{\rm IQH}_{\xi}$) and Laughlin ($\Delta_{\xi}$) states with increasing pairing strength $|U|$ for $2N=6$ and $N_A=3$, $\alpha = 1/8$ (left) and $\alpha = 1/10$ (right).}
\label{fig::Ent_Spec}
\end{figure}

We have computed the PES at small and large $|U|$, starting respectively from the unique and the twofold degenerate ground state. 
We focused on the flux values $\alpha = 1/8$ and $\alpha = 1/10$ for which quasidegeneracies were observed. 
Here, we provide evidence that the large $|U|$ phase shows all features of the Laughlin state from the PES perspective. Additionally, we use the entanglement gap as a witness of the phase transition.
Further details of the PES analysis are provided in~\cite{suppmatCP}.

The nature of the phase at large $|U|$ can be settled by investigating the $N_A =4$ partition. Indeed, it is the smallest partition with multiple (two) BMs in the subsystem $A$, and the PES will reveal their interaction. Fig.~\ref{fig::Ent_Spec}a shows the PES obtained at strong attraction ($|U|=30t$) for $2N=8$ with a particle partition $N_A = 4$ in the sector $S_z^A = 0$. Below the entanglement gap, there are $20$ states ($6,4,6,4$), which is the expected counting for the $\tilde{\nu}=1/2$ bosonic Laughlin state, as known from the generalized exclusion principle~\cite{PhysRevLett.67.937}. This constitutes strong evidence of the emergence of Laughlin physics in this system. Given the large Hilbert space dimension and the smallness of the energy gap we used a truncated Hilbert space (corresponding to the projection onto the space with at least $3$ tightly bound BMs) to obtain the eigenvectors of these two systems before computing their PES. We quantitatively justify this approximation in the supplementary materials~\cite{suppmatCP}.

The entanglement gap $\Delta_\xi$ can be used to monitor the phase transition, providing complementary evidence to the energy spectra. In Fig.~\ref{fig::Ent_Spec}b we plot $\Delta_\xi$ as a function of the interaction strength for $2N=6$ fermions and $N_A=3$, which corresponds to a reasonable computation time. It reveals the transition from an IQH entanglement gap to a Laughlin entanglement gap. Remarkably, the transition point is found for $|U|\simeq 10\,t$, in agreement with the result extracted from the energy spectra (Fig.~\ref{fig::Energy}b).

\begin{figure}[t]
\begin{center}
\includegraphics[width = 86mm]{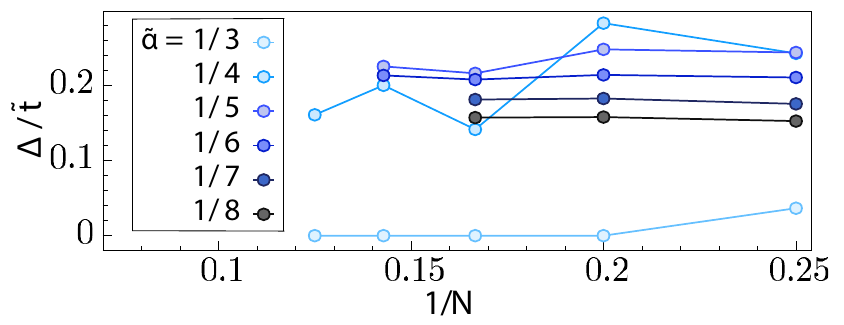}
\caption{Scaling of the energy gap above the twofold quasidegenerate ground states of the hard-core bosonic model Eq.~\eqref{eq:Hamiltonian_bosons} with inverse particle number for various flux densities $\tilde{\alpha}$. There is no quasidegeneracy for  $\tilde{\alpha}>1/4$. The finite size effects are important for $\tilde{\alpha}=1/4$, but become smaller for $\tilde{\alpha}< 1/4$, suggesting a finite gap in the thermodynamic limit.}
\label{fig::Hardcore}
\end{center}
\end{figure}

\paragraph{System size effects ---}
In order to characterize the phase transition from fermionic IQH state to the candidate bosonic phases, it is necessary to reliably extrapolate our results to the thermodynamic limit, which requires large $N$ values. However the spinful fermionic model Eq.~\eqref{eq:Hamiltonian} limits us to systems with only a few pairs of fermions. As a compromise, we focus here on the large $U$ limit, where the model Eq.~\eqref{eq:Hamiltonian} can be mapped onto the following hard-core bosonic Hamiltonian:
\begin{equation}
\label{eq:Hamiltonian_bosons}
\mathcal{H}_{\rm bos} =  - \tilde{t} \sum_{\langle \br,\br' \rangle} \left( e ^{2 i \varphi_{\br\br'}} a_{\br}^{\dagger} a_{\br'} + h.c. \right)
  \end{equation}
where $\tilde{t} = \frac{t^2}{|U|}$ and $a_{\br}^{\dagger}$ (resp. $a_{\br}$) creates (resp. annihilates) a hard-core boson at site $\br$. This Hamitonian allows us to evaluate finite size effects by simulating larger systems as well as to investigate a larger range of $\tilde{\alpha}$. In this limit, we will thus treat the bosonic model Eq.~\eqref{eq:Hamiltonian_bosons} with $N$ bosons at flux density $\tilde{\alpha}= 2 \alpha$ as the analog of the model Eq.~\eqref{eq:Hamiltonian} with $2N$ fermions at flux density $\alpha$. 
The $1/|U|$ scaling of the gap in the large $|U|$ limit (Fig.~\ref{fig::Energy}b) supports this approximation for $|U|/t > 50$. We computed the low energy spectrum of the Hamiltonian Eq.~\eqref{eq:Hamiltonian_bosons} for $1/8\lesssim\tilde{\alpha}\lesssim1/3$ and all numerically accessible particle numbers, extending by up to 4 bosons the computations in~\cite{PhysRevLett.94.086803}.
For $\tilde{\alpha}\lesssim1/4$, we found that the ground state is quasi twofold degenerate.The gap $\Delta$ between the second and third lowest energy states is shown in Fig.~\ref{fig::Hardcore}, and displays a smooth behavior as a function of $N$ for $\tilde{\alpha}<1/4$, which suggests a finite value at the thermodynamic limit. As expected, the finite-size effects decrease as $\tilde{\alpha}$ approaches the continuous limit $\tilde{\alpha}\ll 1$ and they appear to still be important at $\tilde{\alpha}=1/4$. For $\tilde{\alpha}\leq1/4$, the PES analysis yields a number of states below the entanglement gap that is the one expected for the Laughlin state, in agreement with our findings on the fermionic model at even $N_A$.

\paragraph{Conclusion ---}

In this letter, we showed that attractive interactions between opposite spin fermions trigger a phase transition between a $\nu_{\uparrow,\downarrow}=1$ fermionic IQH state and a bosonic Laughlin state in the Harper--Hofstadter model. The transition occurs for a critical interaction strength of the order of the full linewidth of the 1-body spectrum. As this effect is found to occur in the low flux density limit, we also expect a similar phase transition in the continuum, as was suggested by field-theoretical arguments~\cite{PhysRevLett.100.030404}. Our work demonstrates that dynamical fermion pairing, as realized in quantum gas experiments, open novel possibilities for the exploration of topological matter. Combined with the techniques developed for the realization of topological matter with ultracold atoms, this feature might provide a new pathway to create and probe nontrivial topological phases~\cite{Yefsah2016Private}.

\begin{acknowledgements}
The authors thank N.~Regnault for helpful discussions as well as N.~Goldman and A.~Goldsborough for useful comments. T.Y is grateful to Immanuel Bloch for his hospitality at the Max-Planck Institute for Quantum Optics, where part of this work has been done. A.S. is supported by the European Research Council (ERC) under grant no. 636201 WASCOSYS.

\end{acknowledgements}

\bibliography{FQHE_CP.bib}

\clearpage
\newpage

\section{SUPPLEMENTAL MATERIAL}

In this supplementary material, we give additional data supporting the conclusions of the main text. All results described here were obtained via exact diagonalization of the spin--1/2 fermionic Hamitonian Eq.~(1). We focus on flux densities $\alpha = 1/8$ and $\alpha = 1/10$ where a Laughlin-like state was identified in the strongly interacting regime.

\subsection{Quasihole excitations}

\begin{figure}[b!]
\includegraphics{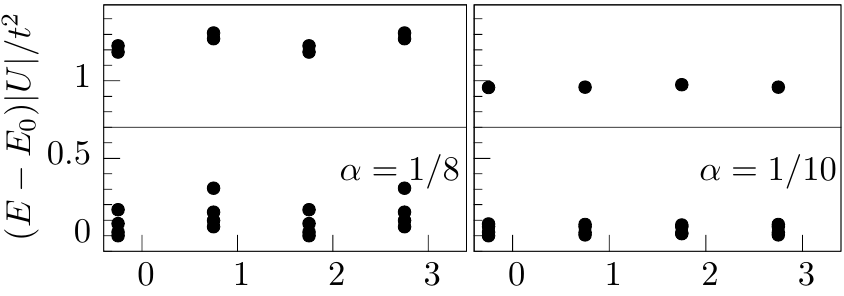}
\caption{Low-energy spectrum of the system with $2N = 6$ fermions at flux densities $\alpha = 1/8$ (\emph{left panel}) and $\alpha = 1/10$ (\emph{right panel}) with an attractive interaction of strength $U = -30t$. The even states under spin inversion have an energy offset $|U|$ and are thus not visible on the graphs. There is a clear many-body gap whose position is indicated by a black line. The number of states below it is $16$ ($4$ per sector) which is the number of quasiholes states expected for the $\tilde{\nu} = 1/2$ Laughlin state with one quasihole.}
\label{fig::qh}
\end{figure}

Quasihole excitations can be created in the system by lowering the number of fermions while keeping the other parameters untouched. Starting from the ground state with $2N = 8$ fermions in the strongly interacting regime ($U = -30t$), we removed $2$ fermions, which corresponds to removing one bosonic molecule (BM). The energy spectrum of this system is shown in Fig.~\ref{fig::qh}. It displays a low-energy manifold of quasi-degenerate states separated by a clear gap from higher energy states. All low-energy states are odd under spin-inversion, as is expected from the fact that they are made of $3$ BMs, and their number ($16$) is the same as the number of low-energy states, the so called quasiholes states, in the bosonic $\tilde{\nu} = 1/2$ Laughlin system with $N = 3$ bosons and $N_{\phi} =8$ flux quanta. The number of states per momentum sector is predicted using the generalized exclusion principle~\cite{PhysRevLett.67.937} and the mapping~\cite{PhysRevB.85.075128} between magnetic and lattice Brillouin zones. 

\subsection{Hilbert space truncation}

\begin{figure}[b!]
\includegraphics{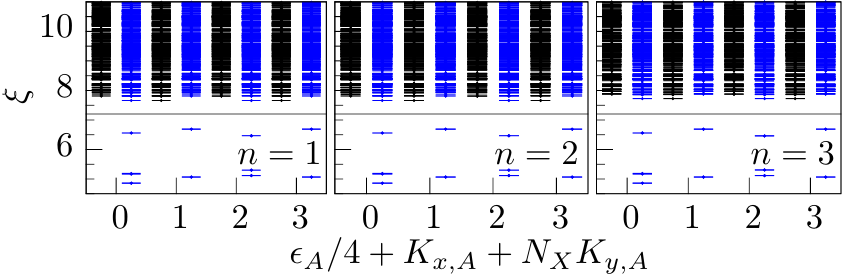}
\caption{PES at $N_A = 4$ and $S_{z,A} = 0$ of the twofold quasidegenerate ground state of the system with $2N = 8$ fermions at $U = -30t$ and flux density $\alpha = 1/8$ obtained with different levels of approximation. The diagonalization of Eq.~1 was performed in the Hilbert space restricted to fermionic configurations with at least $n = 1$ (\emph{left}) $n = 2$ (\emph{middle}) or $n = 3$ (\emph{right}) BMs. The three PES look very similar, and share the same universal features: there is an entanglement gap separating the lowest $20$ $(6, 4, 6, 4)$ states from the rest of the PES, as expected for the bosonic Laughlin $\tilde{\nu} = 1/2$ state. The relative variation of the entanglement gap as the level of approximation is changed from $n = 1$ to $n = 3$ is smaller than $6\%$.}
\label{fig::PES truncation}
\end{figure}

For our numerical calculations, we worked in the Hilbert space defined by all configurations of $2N$ spin--1/2 fermions on $N_s = \frac{N_{\Phi}}{\alpha}$ sites ($N_s = \frac{N}{\alpha}$ at filling fraction $\nu_{\uparrow, \downarrow} = 1$). Due to the large interaction strength $|U|/t \gg 1$, we need to allow the occupation of all bands of the Harper--Hofstadter model, leading to very large Hilbert spaces even for a small number of fermions. Taking all symmetries into account, the largest subspace of the system $2N = 8$, $\alpha = 1/10$ (respectively $\alpha = 1/8$) has a dimension of $1.04\ 10^9$ (resp. $1.7\ 10^8$). Moreover, the very different energy scales of the problem in the strongly interacting $|U|/t \gg 1$ regime add to the computational burden. Roughly speaking, the gap $\Delta$ above the twofold ground state is of the order of $\frac{t^2}{|U|}$, while the entire spectrum has an amplitude $N|U|$. The small ratio $\frac{t^2}{NU^2}$ between these two quantities is thus particularly unfavorable, and leads to very long convergence times since it controls the convergence of the Lanczos algorithm used here.

Given these computational constraints, it is advantageous to diagonalize the Hamiltonian Eq.~(1) in a truncated basis. To build the truncated Hilbert space, we discard the fermionic configurations with (strictly) less than $n$ BMs. For this purpose, we consider that two fermions with opposite spins at the same lattice position constitute a BM.  At large $|U|/t$, the configurations with the smallest numbers of BMs have the smallest coupling elements in the Hamiltonian matrix, and suppressing them will only lead to minor modifications of the ground state. Note that the constraint $n = N$ is equivalent to a Hilbert space of hard-core bosons, while $n = 0$ corresponds to the full fermionic Hilbert space. 

To compute the particle entanglement spectra (PES) at $2N = 8$ (Fig.~ 3a of the main text), we restricted the Hilbert space to the configurations with at least $n = 3$ BMs. We have quantitatively checked the validity of this approximation by computing the weight of the $n = 1$ twofold ground state on the $n = 2, 3, 4$ Hilbert spaces. For $\alpha = 1/10$ (resp. $\alpha = 1/8$), these weights are $0.99998$, $0.99859$, $0.93944$ (resp. $0.99998$, $0.99866$, $0.94083$) at $U = -30t$ where the PES calculations were performed. For the same interaction strength and $2N = 6$ where the ground state can be computed without any restriction ($n = 0$), its respective weights on the $n =1, 2, 3$ Hilbert spaces are $0.99999$, $0.99928$ and $0.95420$ for $\alpha = 1/10$ (respectively $0.99999$, $0.99932$ and $0.95529$ for $\alpha = 1/8$). These very high weights show that the ground state is essentially captured by the $n = N - 1$ Hilbert space at $U = -30t$ thus validating the use of a truncated Hilbert space to compute the PES.

Additionally, we computed the PES of the twofold ground state at $2N = 8$, $\alpha = 1/8$ and $U = -30t$ for various degrees of truncation. Fig.~\ref{fig::PES truncation} shows the PES at $N_A = 4$ starting from the ground state obtained while keeping configurations with at least $n = 1, 2$ and $n=3$ BMs. While these PES are numerically different, it is almost impossible to distinguish them, and the relative variation of the entanglement gap between $n = 1$ and $n = 3$ is less than $6\%$. This further validates the use of the approximation in Fig.~3a of the main text.

\subsection{Particle entanglement spectra}

\begin{figure}
\includegraphics{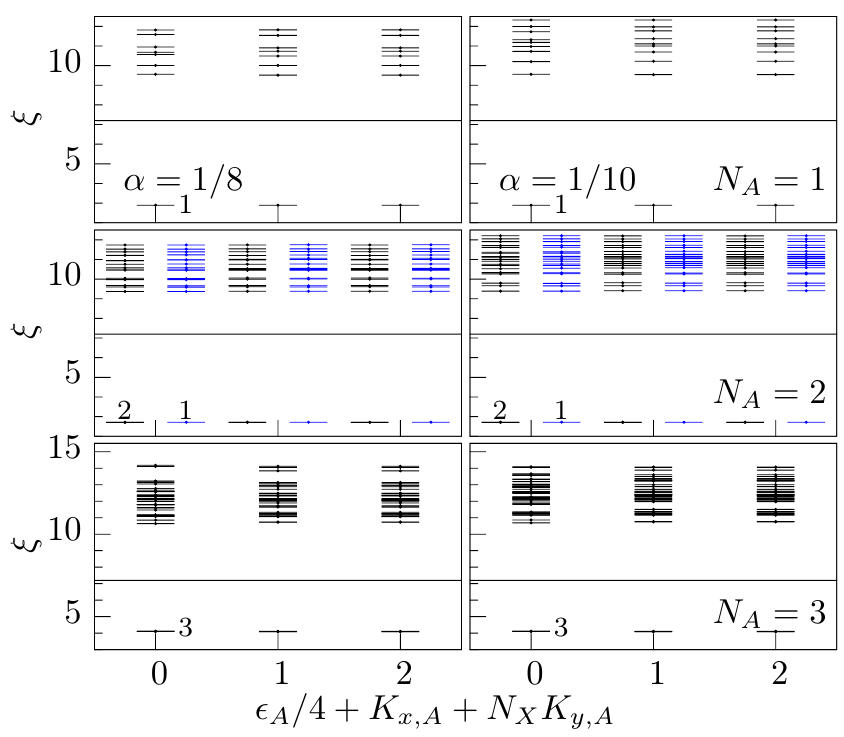}
\caption{PES of the unique ground state of the system with $2N = 6$ fermions at $U = -t$ and flux density $\alpha = 1/8$(\emph{left panel}) and $\alpha = 1/10$ (\emph{right panel}). For the sake of clarity, we only show the lowest entanglement levels. The particle partition is characterized by $N_A = 1, 2, 3$ and respectively $S_{z,A} = 1, 0, 1$. When $S_{z,A} = 0$, the odd (resp. even) states under spin inversion are represented in black (resp. blue). By convention, we set $\epsilon_A = 0$ when spin inversion is not a symmetry of the system ($S_{zA} \neq 0$). The black line materializes the position of the entanglement gap. The number of states below it is the same in each momentum sector and is indicated next to each state.}
\label{fig::PES IQH}
\end{figure}

To complement the PES analysis of the main text, we give a detailed account of all the PES features observed in the weakly and strongly interacting regimes. They are in full agreement with the conclusions of the main text.

Fig.~\ref{fig::PES IQH} shows the PES of the unique ground state obtained in the weakly interacting regime ($U = -t$) with $2N = 6$, $N_A = 1, 2, 3$. We observe a clear entanglement gap for all values of $N_A$, and the number of states below it is given by the number of ways to distribute $N_A$ fermions in the lowest band of the Harper--Hofstadter model, confirming the integer quantum Hall nature of the ground state. In a given entanglement spectrum sector defined by $(N_A,S_{z,A})$, it is indeed ${N_{\Phi} \choose N_{A,\uparrow}}{N_{\Phi} \choose N_{A,\downarrow}}$, where $N_{A,\uparrow} =  N_A/2 + S_{z,A}$ (resp. $N_{A,\downarrow}= N_A/2 - S_{z,A}$) is the number of spin up (resp. down) fermions in subsystem A.

\begin{figure}
\includegraphics{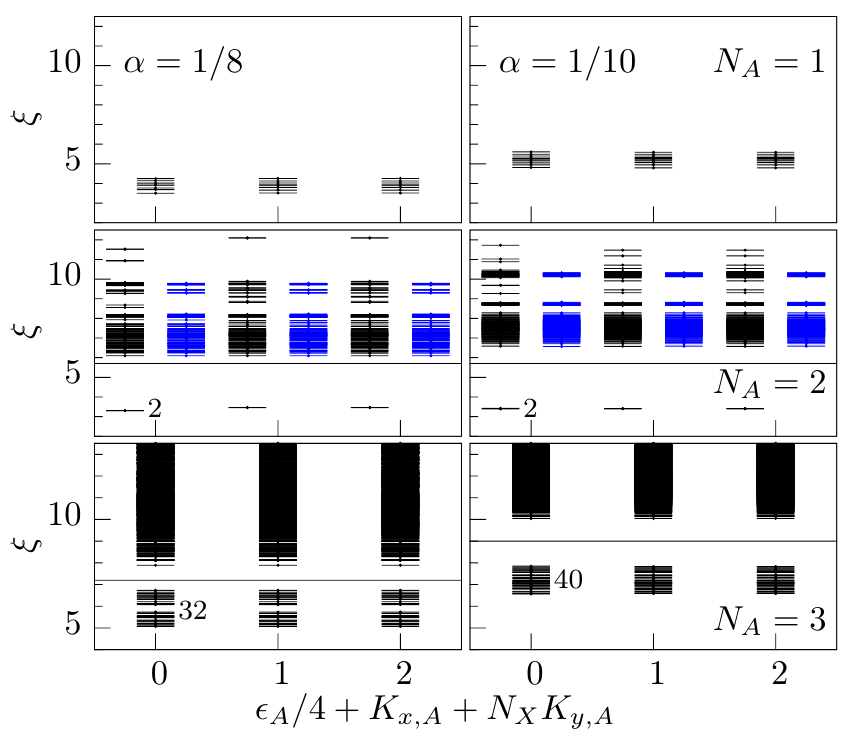}
\caption{PES of the twofold quasidegenerate ground state of the system with $2N = 6$ fermions at $U = -30t$ and flux density $\alpha = 1/8$ (\emph{left panel}) and $\alpha = 1/10$ (\emph{right panel}). For the sake of clarity, we only show the lowest entanglement levels. The particle partition is characterized by $N_A = 1, 2, 3$ and respectively $S_{z,A} = 1, 0, 1$. When $S_{z,A} = 0$, the odd (resp. even) states under spin inversion are represented in black (resp. blue). By convention, we set $\epsilon_A = 0$ when spin inversion is not a symmetry of the system ($S_{zA} \neq 0$). The entanglement levels $N_A = 1$ PES all have the same order of magnitude. The black line materializes the position of the entanglement gap in the $N_A = 2, 3$ PES. The number of states below it is the same in each momentum sector and is indicated next to each state.}
\label{fig::PES FQH}
\end{figure}

In contrast, the PES obtained in the strongly attractive regime display different behaviors depending on the parity of $N_A$, due to the paired nature of the state.
Fig.~\ref{fig::PES FQH} shows the PES of the twofold almost degenerate ground state obtained in the strongly interacting regime ($U = -30t$) with $2N = 6$, $N_A = 1, 2, 3$. 

Let us first focus on the even values of $N_A$. In this case, we always observe an entanglement gap, and the number of states below it is equal to the number of quasihole states in the bosonic system with twice the flux. These low energy levels appear only in the sector $S_{z,A}=0$ and have a parity $\epsilon_A=(-1)^{N_A/2}$ under spin inversion. For $N_A=2$, their number $2N_{\Phi}$ is the number of single-particle states in the lowest band of the BM model.
For $2N = 6$ and $N_A=2$ (Fig.~\ref{fig::PES FQH}), there are indeed $6$ states below the entanglement gap.
This confirms the formation of BMs in the lowest band at large $|U|$. The $N_A = 4$ PES (shown in the main text for $2N = 8$) probes their many-body behavior, and has the same counting as the $\tilde{\nu} = 1/2$ Laughlin state.

Let us now consider the odd values of $N_A$. For $N_A=1$, the reduced density matrix is full rank and all its eigenvalues have the same order of magnitude, as is clearly visible in Fig.~\ref{fig::PES FQH}. This demonstrates that the large $|U|$ phase is completely featureless in the fermionic language. 
For $N_A = 3$, we observe an entanglement gap $\Delta_{\xi}$.
As can be seen in Fig.~\ref{fig::PES FQH}, the number of states below it is respectively $3 \times 32$ and $3 \times 40$ in the $2N = 6$, $\alpha = 1/8$ and $\alpha = 1/10$ systems (or generically $N \times 4 \alpha^{-1}$).
The low entanglement levels appear in the $S_{z,A} = \pm 1/2$ sector and can be interpreted as the low energy spectrum of a BM and a free fermion. In this picture, the BM occupies one of the $2N_{\phi} = 2N$ Landau orbitals of the BM lowest band and is thus delocalized on $\frac{N_{\rm s}}{N}$ sites. Due to the Pauli principle, the fermion can access any of the $(N - 1) N_{\rm s}/N $ remaining sites, leading to the observed counting.

\end{document}